\documentclass[aps,prl,twocolumn,showpacs,superscriptaddress,groupedaddress,nofootinbib,floatfix]{revtex4-1}

\usepackage{graphicx,psfrag}
\usepackage{mathrsfs}
\usepackage{amsmath,amsfonts,amssymb}
\usepackage{multirow,hyperref}
\usepackage{comment}
\usepackage{enumerate}
\usepackage{mathrsfs}
\usepackage{pifont}
\usepackage[normalem]{ulem}
\usepackage{booktabs}

\newcommand{\be}{\begin{equation}}
\newcommand{\ee}{\end{equation}}
\newcommand{\bea}{\begin{eqnarray}}
\newcommand{\eea}{\end{eqnarray}}
\newcommand{\bel}{\begin{align}}
\newcommand{\eel}{\end{align}}

\def\non{\nonumber}                     

\def\e{{\rm e}}
\def\i{{\rm i}}

\def\GMc2{{\rm G M_{\odot} c^{-2}}}

\def\mo{\hat{\omega}}
\def\ct{\tilde{c}} 
 
\def\nt{\tilde{n}} 
\def\dt{\tilde{d}}

\usepackage{color}

\begin{document}

\title{Closed-form tidal approximants for binary neutron star
  gravitational waveforms\\
constructed from high-resolution numerical relativity simulations}

\author{Tim Dietrich$^1$}
\author{Sebastiano Bernuzzi$^{2,3}$}
\author{Wolfgang Tichy$^4$}

\affiliation{${}^1$Max Planck Institute for Gravitational Physics (Albert Einstein Institute), Am M\"uhlenberg 1, Potsdam 14476, Germany}
\affiliation{${}^2$Department of Mathematical, Physical and Computer Sciences, University of Parma, I-43124 Parma, Italy}  
\affiliation{${}^3$Istituto Nazionale di Fisica Nucleare, Sezione Milano Bicocca, gruppo collegato di Parma, I-43124 Parma, Italy}  
\affiliation{${}^4$Department of Physics, Florida Atlantic University, Boca Raton, FL 33431 USA}

\date{\today}

\begin{abstract}
We construct closed-form gravitational waveforms (GWs) with tidal
effects for the coalescence of binary neutron stars. 
The method relies on a new set of eccentricity-reduced and
high-resolution numerical relativity (NR) simulations and is composed of
three steps. 
First, tidal contributions to the GW phase are extracted from
the time-domain NR data. Second, those 
contributions are employed to fix high-order coefficients in an
effective and resummed post-Newtonian expression.
Third, frequency-domain tidal approximants are built using the
stationary phase approximation.  
Our tidal approximants are valid from the low frequencies to the
strong-field regime. They can be analytically added
to any binary black hole GW model to obtain a binary
neutron star waveform, either in the time or in the frequency domain. 
This work provides simple, flexible, and accurate models ready to be used in
both searches and parameter estimation of binary neutron star events. 
\end{abstract}


\maketitle

%% ______________________________________________________________
\section{Introduction}

The 2015 detections of gravitational waves (GWs) of 
merging binary black holes (BBHs)~\cite{Abbott:2016blz,Abbott:2016nmj} 
have initiated a new observational era in astronomy and fundamental
physics.   
In the coming years, ground-based advanced interferometers will
reach design sensitivity and observe the coalescence and
merger of binary neutron stars (BNSs)~\cite{Abbott:2016ymx}. 
These observations will have a unique potential to probe the fundamental
physics of NSs and to connect high-energy astrophysical phenomena 
with their strong-gravity engines. 
The main examples are the possibility to constrain the equation of state 
(EOS) of the cold ultradense matter in
NS interiors, e.g.~\cite{DelPozzo:2013ala}, and
the possibility to show the unequivocal connection between 
electromagnetic signals, e.g.~short gamma ray bursts~\cite{Paczynski:1986px} or
kilonovae~\cite{Tanvir:2013pia}, 
with the collision of two compact objects.

A key open problem for GW astronomy with BNS sources is the
availability of faithful waveform models that capture the
strong-gravity and tidally dominated regime of the late-inspiral and
merger. State-of-the-art tidal waveform models have been developed in 
\cite{Bernuzzi:2014owa,Hinderer:2016eia} and are based on the 
effective-one-body (EOB) description of the general-relativistic
two-body problem \cite{Buonanno:1998gg,Damour:2009wj}. 
That approach proved to be very powerful but has also limitations.
EOB waveforms cannot be efficiently evaluated, hence they cannot be
directly used for GW searches or parameter estimation. Fast
representations of EOB can be build using reduced-order-modeling
techniques~\cite{Lackey:2016krb}, but they require extra efforts and
introduce further uncertainties.  
Additionally, the currently published tidal EOB models 
neither include spin effects nor are tested against spinning NR
simulations \cite{Bernuzzi:2013rza}. 
Recent work also showed that the current EOB models are not 
uniformly accurate on the binary parameter space that has been
simulated in Ref.~\cite{Hotokezaka:2015xka,Dietrich:2017feu}. Thus, 
modeling techniques complementary
to EOB, see e.g.~\cite{Lackey:2013axa,Barkett:2015wia}, are needed
especially because post-Newtonian (PN) approximants
fail towards merger and introduce systematic uncertainties in GW
parameter estimation
\cite{Bernuzzi:2012ci,Favata:2013rwa,Wade:2014vqa}. 

In this work, we construct for the first time closed-form (analytical) approximants to the
tidal GW phase directly employing numerical relativity (NR)
simulations. Simple time and frequency domain approximants are build
from a set of error-controlled BNS merger simulations. 
Our method is inspired by some ideas used in the modeling of BBH's 
GWs. In particular, it makes direct use of NR data as in the Phenom
approach~\cite{Khan:2015jqa} and employs resummed PN expressions 
as in the EOB approach. 

%% ______________________________________________________________
\section{Eccentricity-reduced and high-resolution NR simulations}
For this work, we simulated nine BNS configurations in
general relativity. We simulated equal-mass BNSs both irrotational
and with spins (anti) aligned to the orbital angular momentum. 
Three different parameterized EOSs (MS1b, H4,
SLy)~\cite{Read:2008iy} are employed to span a large range of tidal 
parameters (see below). The binary gravitational mass
is $M=M_A+M_B\sim2.7$, where $A,B$ label the NSs and $M_A$ is the mass
of star $A$ in isolation. Spin magnitudes are in the range
$\chi_A=\chi_B\sim[-0.1,+0.15]$, where $\chi_A=S_A/M_A^2$ is the mass-rescaled
dimensionless spin. 
We use the numerical methods implemented in the pseudospectral initial
data SGRID code \cite{Tichy:2012rp} and in the 3+1
adaptive-mesh-refinement evolution BAM code
\cite{Thierfelder:2011yi}. Key technical points are 
the use of the Z4c formulation of general relativity and of a 
high-order scheme for the hydrodynamics \cite{Hilditch:2012fp,Bernuzzi:2016pie}.
See \cite{supplmat} for further details. Note that we employ geometric
units $G=c=M_\odot=1$. 

These new simulations significantly improve the waveform's
quality over previous ones.
Low-eccentricity initial data were generated following 
Ref.~\cite{Dietrich:2015pxa}; our BNSs have $e \sim 10^{-3}$.
Each BNS is evolved using four to five grid resolutions making a total 
of 37 runs. 
The NSs are resolved with smallest grid spacings in the
range $dx=0.291-0.059$ in each direction. 
These are the largest BNS simulations performed with the BAM code so far 
and utilized $\sim 25$ million CPU hours on various
high-performance-computing clusters. 
Numerical uncertainties are estimated from convergence tests and a
detailed error budget has been computed.
Our waveforms have maximal errors at merger, accumulated over
$\sim12$ orbits, of $\sim0.5-1.5$ radians, depending on the
particular configuration~\cite{supplmat}. 
The waveforms are publicly available under \texttt{www.computational-relativity.org}; see
Refs.~\cite{Dietrich:2018upm,Dietrich:2018phi} for more details.

%% ______________________________________________________________
\section{Extraction of tidal contributions}

Spin and tidal effects in the phase of the complex GW 
$h(t)=A(t) \e^{-\i\phi(t)}$ are parametrized to leading PN order, 
respectively, by the effective spin 
\begin{equation} \label{eq:chi}
  \chi_{\rm eff} = X_A \chi_A + X_B \chi_B - \frac{38}{113} X_A X_B(\chi_A + \chi_B)
\end{equation}
describing the spin-orbit (SO) interaction \cite{Blanchet:2013haa}, and by
an effective tidal coupling constant \cite{Damour:2009wj,Damour:2012yf} 
\begin{equation}\label{eq:kappa}
\kappa^T_{\rm eff} = \frac{2}{13} \left[ 
\left(1+12\frac{X_B}{X_A}\right)\left(\frac{X_A}{C_A}\right)^5 k^A_2 + 
 (A \leftrightarrow B) 
\right]  \ ,
\end{equation}
where $k^A_2$ is the quadrupolar Love number describing the static
quadrupolar deformation of one body in the gravitoelectric field of
the companion, $X_A=M_A/M$, and $C_A$ is the compactness of star $A$.
$\kappa^T_{\rm eff}$ is defined here for the first time but it based
on the expressions for generic mass ratio phasing in \cite{Damour:2012yf}.
For equal mass systems $\kappa^T_{\rm eff}$ is identical to the 
dimensionless tidal coupling constant $\kappa^T_2$ defined 
in~\cite{Damour:2009wj,Bernuzzi:2014kca}. 

\begin{figure}[t]
   \includegraphics[width=0.48\textwidth]{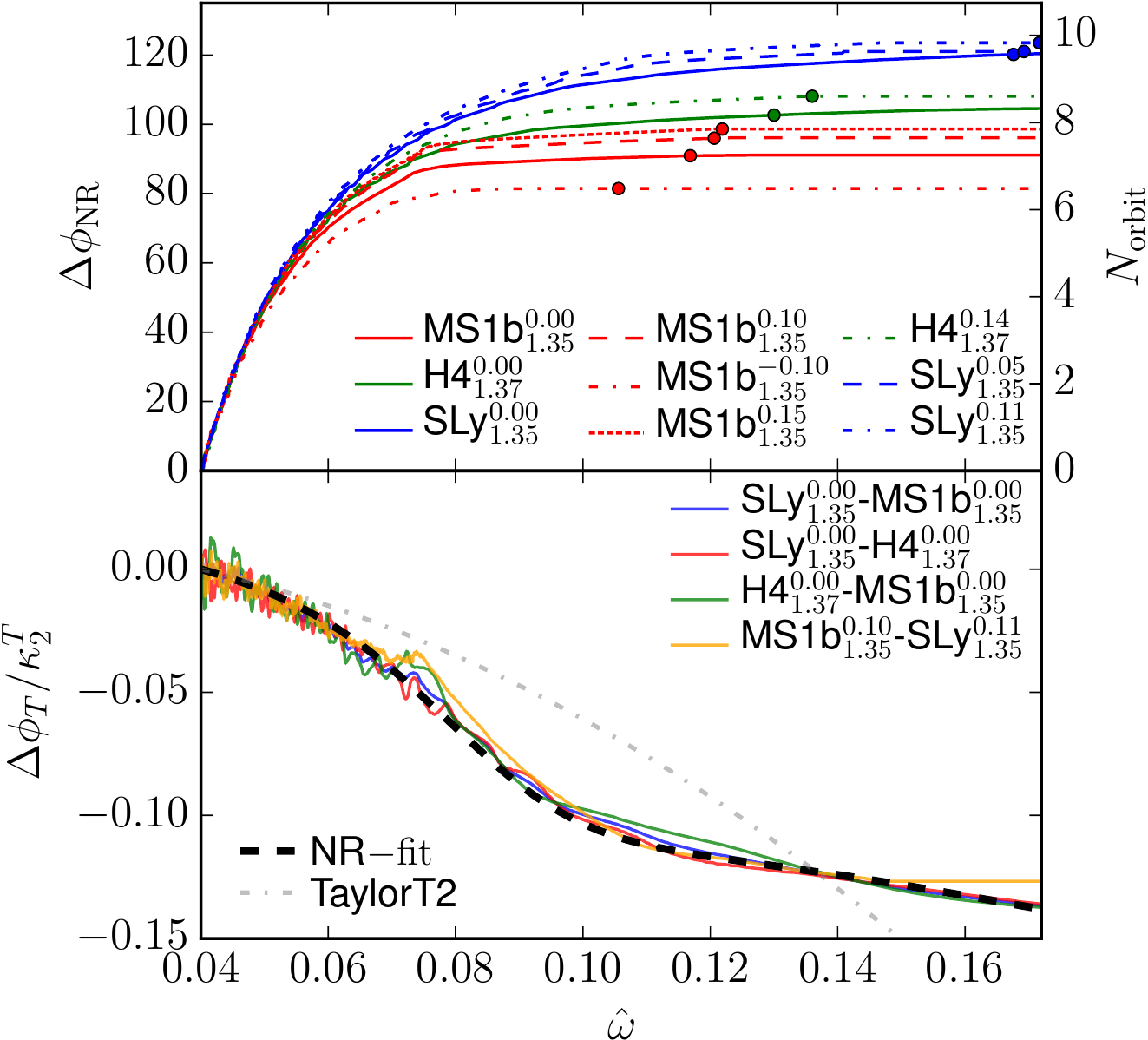}
   \caption{Phase as a function of the GW frequency from NR simulations.
     The simulations are labeled as EOS$_{M_A}^{\chi_A}$.
     Top: Total phase / number of cycles accumulated within frequency interval
     $\mo\in[0.04,0.17]$ for different BNSs.
     Markers indicate the merger (peak of the GW's amplitude)
     of the particular simulation for the highest revolved simulation.   
     Bottom: Pairwise phase differences equivalent to the tidal phase
     $\phi_T/ \kappa_{\rm eff}^T$; note the spin independence.}
   \label{fig:Phiomega}
\end{figure}

In order to separate the tidal phase, we work with the phase as a
function of the dimensionless GW frequency 
$\mo = M \partial_t {\phi}(t)$ and use the PN {\it ansatz},
\begin{equation}\label{eq:phi_omg}
 \phi(\mo) \approx \phi_0 (\mo) + \phi_{SO}(\mo) + \phi_T(\mo) \ ,
\end{equation}
where $\phi_0$ denotes the nonspinning black hole (or 
point particle) phase evolution. The SO contribution is 
$\phi_{\rm SO}\propto\chi_{\rm eff}$ at leading 1.5PN order and it is
currently known up to 3.5PN order. 
For simplicity, we neglect
spin-spin interactions; they are subdominant contributions and poorly 
resolved in our simulations \cite{Dietrich:2016lyp}.
Tidal contributions enter the
phasing at 5PN. The currently known next-to-leading-order 
PN expression of the tidal contribution (TaylorT2
approximant)~\cite{Wade:2014vqa} reads
\begin{equation}\label{eq:T2} 
\phi_T^{\rm T2} = - \kappa_{\rm eff}^T \frac{c_{\rm Newt}
  x^{5/2}}{X_A X_B}  (1 + c_1 x )  \ ,
\end{equation}
with $x(\mo)=(\mo/2)^{2/3}$, where $\mo/2$ is the orbital frequency,   
and $c_{\rm Newt} = -13/8$, $c_1 = 1817/364$ (value for equal mass 
case). 
Using Eq.~\eqref{eq:phi_omg} the nonperturbative SO and tidal
contributions can be extracted by linearly combining pairs of simulation 
data with different parameters, as detailed in
\cite{Bernuzzi:2013rza,Dietrich:2016lyp,supplmat}. 
The top of Fig.~\ref{fig:Phiomega} shows the total phase accumulated over
simulations. The bottom shows the phase differences divided by the
differences in $\kappa^T_\text{eff}$ for several simulation pairs
\cite{supplmat}, denoted by $\Delta \phi_T/\Delta\kappa^T_\text{eff}$.
According to Eqs.~\eqref{eq:phi_omg} and \eqref{eq:T2} 
$\Delta \phi_T/\Delta\kappa^T_\text{eff} \approx\phi_T/\kappa^T_\text{eff}$.
For comparison we also show $\phi_T/\kappa^T_\text{eff}$ of our fit and
of TaylorT2. We find that the leading-order EOS effect is captured well by
$\kappa^T_\text{eff}$ and the residual dependency on the EOS, related to
multipolar tidal coefficients with $\ell>2$, is negligible. Most
importantly, tidal interactions decouple from spin interactions for
the spin values explored by NR data and at level of the NR
uncertainties. This fact allows us to  
construct spinning BNSs using binary black hole baseline waveforms and
adding the tidal contribution. 
Further, Fig.~\ref{fig:Phiomega} indicates that the TaylorT2
approximant does {\it not} capture the phase evolution in the strong
field region, failing for $\mo \gtrsim 0.06$, which is approximately
the contact frequency~\cite{Bernuzzi:2012ci}.

%% ______________________________________________________________
\section{Time-domain tidal approximant}

A closed-form expression for $\phi_T$ is
obtained using the fitting formula
\begin{small}
\begin{align}\label{eq:fitT} 
\phi_T &= 
-\kappa_{\rm eff}^T\frac{c_{\rm Newt}}{X_A X_B} x^{5/2} \ \times \\%\nu  
& \frac{1  + n_1 x + n_{3/2} x^{3/2} + n_2 x^{2} + n_{5/2} x^{5/2}+n_3 x^3}
      {1+ d_1 x + d_{3/2} x^{3/2}} \non
\end{align}
\end{small}
Demanding that Eq.~\eqref{eq:fitT} reproduces Eq.~\eqref{eq:T2} in a low
frequency expansion, we set $d_1=(n_1-c_1)$. The other coefficients
are fit to NR data.
Note that for simplicity Eq.~\eqref{eq:fitT} does not contain tidal terms
corresponding to higher multipoles \cite{Damour:2012yf}, and the
dependency from $X_{A,B}$ of the higher effective PN terms is
ignored. This is justified since we seek an effective expression
of the phase; the coefficients of the latter could be further improved 
using more simulations with various mass ratios. 

The fit is performed on a dataset spanning the interval
$\mo\in[0,0.17]$. Eq.~\eqref{eq:T2} is used for
$\mo \leq 0.0074$, while the tidal EOB waveforms
of~\cite{Bernuzzi:2014owa} are used for $\mo \geq \in[0.0074,0.04]$. 
The datasets are connected such that phase differences 
near the interval boundaries are minimal. 
We interpolate the data on a
grid consisting of $10000$, $5000$, $500$ points in the three
intervals, respectively. 
Although the final fit depends only weakly on the exact
number of points of the interpolating grid, using more points at lower
frequencies helps constraining the fit in that regime.
Our approximant is defined by Eq.~\eqref{eq:fitT} with the fitting coefficients 
$(n_1,n_{3/2},n_2,n_{5/2},n_3)= ($-17.941$,$57.983$,$-298.876$,$964.192$,$-936.844$)$, 
and $d_{3/2}=43.446$.

\begin{figure}[t]
  \includegraphics[width=0.48\textwidth]{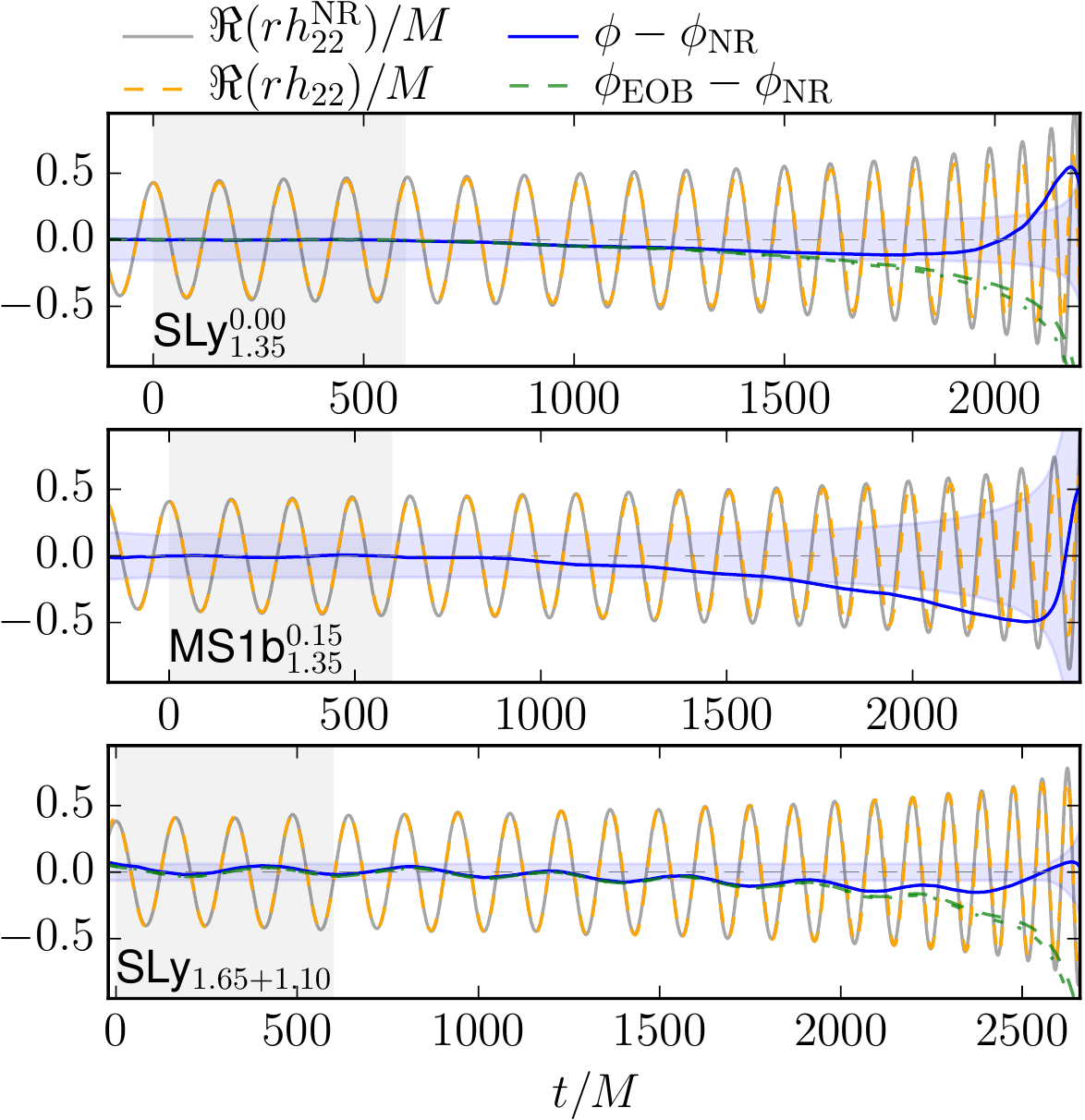}
  \caption{Comparison of NR simulations with model waveforms obtained 
  following Eq.~\eqref{eq:phi_omg}. 
  The panels show the real part of the GW signals (NR data -- gray,
  tidal approximant -- orange).
  We also include the phase between the NR data with respect to 
  our tidal approximant Eq.~\eqref{eq:fitT}, to Taylor T2 tidal approximant 
  Eq.~\eqref{eq:T2} (cyan), and for some cases to 
  EOB (green dashed~\cite{Hinderer:2016eia}, 
  green dot dashed~\cite{Bernuzzi:2014owa}.
  We also indicate the estimated uncertainty of the 
  NR data (blue shaded) and the alignment region (gray shaded).
  Simulations use the same notation as in Fig.~\ref{fig:Phiomega} except 
  for the unequal mass case of~\cite{Dietrich:2017feu} with EOS$_{M_A+M_B}$.}
  \label{fig:timedomain}
\end{figure} 

A time-domain approximant of a BNS configuration
is computed by prescribing $\kappa_{\rm eff}^T$ and 
adding Eq.~\eqref{eq:fitT} to a BBH baseline, i.e.~to $\phi_0$. 
To construct a generic spin-aligned BNS configurations with 
spin $\chi_{\rm eff}$ we use a BBH waveform that includes already the spin contribution, 
i.e.~use as baseline the GW phase of a BBH setup which has the 
same dimensionless spin as the BNS configuration which we are going to 
model.
The time-domain phasing is then calculated by numerically integrating
$t = \int \text{d} \phi /\mo(\phi)$ 
in order to obtain a parametric representation of the tidal phase. 
We stop the integration once $ \phi (\mo)$ reaches its maximum.

Examples of such constructed waveforms are reported in
Fig.~\ref{fig:timedomain}. There, we use the nonspinning BBH waveforms 
from the SXS-database \cite{SpEC,Mroue:2013xna}, in particular setup 66 for
the equal mass cases and setup 7 for the $q = X_A/X_B=1.5$
configuration. 
In order to compare with the BNS configuration with $\chi_{\rm eff}= +0.123$ we 
add to the nonspinning NR BBH curve  
the spin-orbit contributions given in~Eq.~(417)
of~\cite{Blanchet:2013haa}. In general a spinning binary black 
hole baseline should be used. 

In most cases our new waveforms are compatible with the
NR data within the estimated uncertainties.
The proposed tidal approximant remains accurate also for longer waveforms. 
Phase differences with respect to hybrid tidal EOB-NR waveforms
and accumulated over the last 300 orbits before merger are of the order 
of $\sim 1$ rad, see~\cite{supplmat}.
In the nonspinning cases, our results can be directly compared to the
tidal EOB waveforms of \cite{Bernuzzi:2014owa,Hinderer:2016eia} [see
green lines in Fig.~\ref{fig:timedomain}]; comparable performances are observed 
in spite of the simplicity of our model.
The fit gives a good prediction also for the unequal mass 
case, although only the leading-order effect of the mass ratio 
is taken into account, see Eq.~\eqref{eq:kappa}.
Also, while we use NR data up to $\hat{\omega}=0.17$, the model
remains accurate for BNSs with smaller $\kappa_{\rm eff}^T$ that merge at higher
frequencies.
Let us stress that the model performances are independent of the BBH
baseline, provided the latter is a faithful representation
of BBH waveforms.

%% ______________________________________________________________
\section{Frequency-domain tidal approximant}
In the frequency domain, 
$\tilde{h}(f)=f^{-7/6}\tilde{A}(f)\e^{-\i\Psi(f)}$. The expression of the
tidal phase is computed using the stationary phase approximation (SPA)
\cite{Damour:2012yf}  
\begin{equation}\label{eq:PsiT_eq}
  \frac{d^2 \Psi_T^{\rm SPA}}{d \omega_f^2} =
  \frac{Q_\omega(\omega_f)}{\omega_f^{2}} \ ,
\end{equation}
where $\omega_f$ is the Fourier domain circular frequency
$\omega_f=2\pi M f$, and  $Q_\omega (\omega) = d \phi/d\log{\omega}$. 
The integration of Eq.~\eqref{eq:PsiT_eq} with~\eqref{eq:fitT} is
performed numerically; the constants of integration are fixed by
demanding continuity with the TaylorF2$_{\rm 1PN}$ in the limit $f\to0$. The
resulting expression $\Psi_T^{\rm NR}$ can be approximated 
by a Pad\'e function:
\begin{align}\label{eq:PsiT_pade} 
\Psi_T^{\rm NRP} & = -\kappa_{\rm eff}^T\frac{\ct_{\rm Newt}}{X_AX_B}
x^{5/2}\ \times \\
& \frac{1  + \nt_1 x + \nt_{3/2} x^{3/2} + \nt_2 x^{2} + \nt_{5/2} x^{5/2}}
      {1+ \dt_1 x + \dt_{3/2} x^{3/2}} \non
\end{align}
with 
$\ct_{\rm Newt} = 39/16$ and 
$\dt_1 = \nt_1 - 3115 /1248$, the other parameters read: 
$(\nt_1,\nt_{3/2},\nt_{2},\nt_{5/2})=($-17.428$,$31.867$,$-26.414$,$62.362$)$
and $\dt_{3/2}=36.089$.
\begin{figure}[t]
   \includegraphics[width=0.48\textwidth]{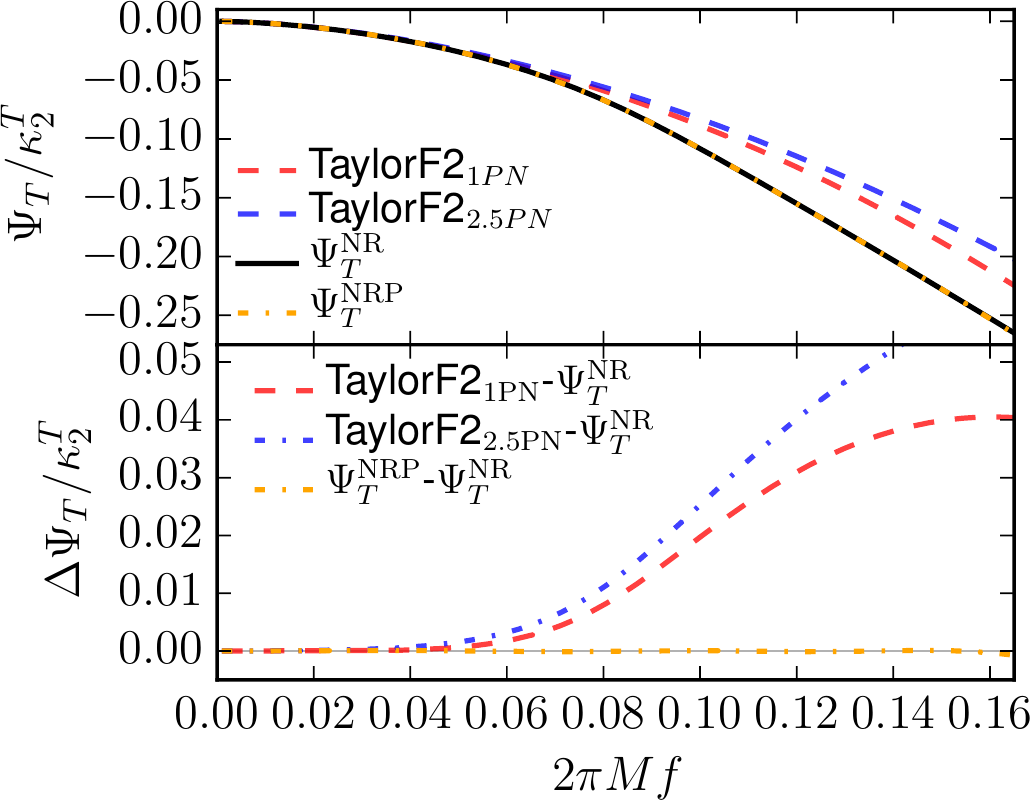}
  \caption{Frequency-domain tidal approximants. 
           Top panel shows $\Psi_T/\kappa_{\rm eff}^T$ as given by the 
           TaylorF2$_{\rm 1PN}$, TaylorF2$_{\rm 2.5PN}$~\cite{Damour:2012yf},
           Eq.~\eqref{eq:PsiT_eq}, and Eq.~\eqref{eq:PsiT_pade}. 
           Bottom panel: Difference between the frequency-domain representations.
           }
  \label{fig:freqdomain}
\end{figure} 

Figure~\ref{fig:freqdomain} compares the obtained tidal approximants 
$\Psi_T^{\rm NR}, \Psi_T^{\rm NRP}$ with the 
TaylorF2$_{\rm 1PN}$ and the 2.5PN approximants given in~\cite{Damour:2012yf} 
(TaylorF2$_{\rm 2.5PN}$).
Because of the construction of Eq.~\eqref{eq:PsiT_pade} the low frequency 
behavior of TaylorF2 is recovered. At higher frequencies PN expressions 
predict smaller tidal effects than $\Psi_T^{\rm NR}$. Considering the
accuracy of  $\Psi_T^{\rm NRP}$, the Pad\'e fit recovers $\Psi_T^{\rm NR}$
with fractional errors $ \lesssim 1\%$.

To further test the performance of the proposed frequency-domain model
we compute the unfaithfulness ($\bar{F} = 1-F$, one minus
faithfulness) which is the mismatch for the fixed intrinsic 
binary parameters with respect to tidal EOB
waveforms starting at $\sim 25$Hz and hybridized with 
NR simulations~\cite{supplmat}. 
The unfaithfulness quantifies the loss in the signal-to-noise ratio
(squared) due to the inaccuracies in the signal modeling.
The typical maximum value used in the GW searches is $\bar{F}\le 0.03$,
which roughly corresponds to $\lesssim 10$\% loss in the number of
events (assuming that they are uniformly distributed).

\begin{figure}[t]
   \includegraphics[width=0.48\textwidth]{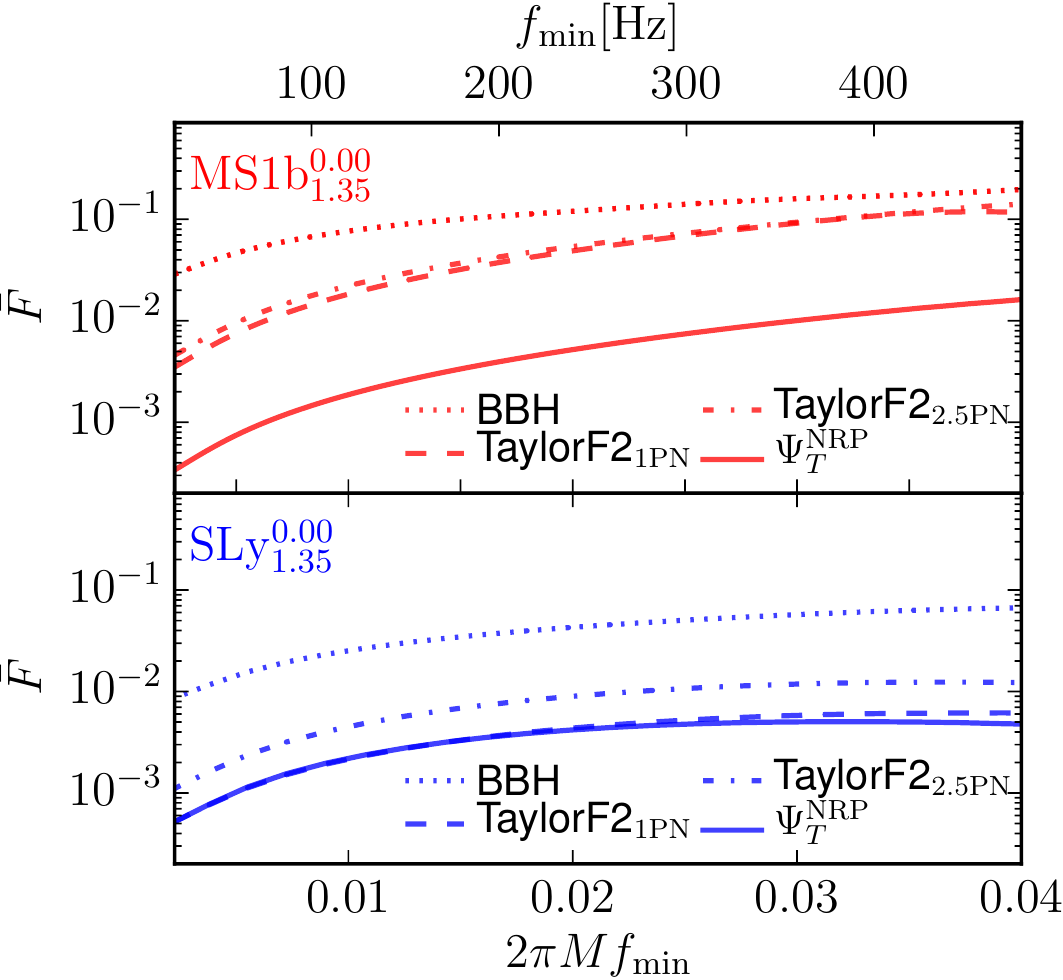}
  \caption{Unfaithfulness of different approximants with respect to
    hybridized EOB-NR waveforms for MS1b$_{1.35}^{0.00}$ (top) 
    and SLy$^{0.00}_{1.35}$ (bottom).
    The unfaithfulness is computed within the interval $M[f_{\rm min},
      f_{\rm max}]\sim[0.0022,0.04]/2\pi$, i.e. $f_{\rm min}$
    varies between $27$ and $480$~Hz. $f_{\rm max}$ is set to the
    merger frequency of the highest resolved simulation 
    (1398~Hz for MS1b$_{1.35}^{0.00}$ and 2005~Hz for SLy$^{0.00}_{1.35}$). 
    As BBH baseline for $\Psi_T^{\rm NRP}$ we use a nonspinning equal-mass 
    EOB waveform~\cite{Nagar:2015xqa}.}
  \label{fig:mismatches}
\end{figure}

Figure~\ref{fig:mismatches} shows $\bar{F}$ for different approximants
and varying the minimum frequency in the overlap interval from
$Mf_{\rm min}\sim[0.0022,0.04]/2\pi$, 
i.e.~from $\sim 27$~Hz to the NR regime ($\sim480$~Hz). 
Tidal approximants have significant
mismatches with respect to BBH ones already for $M f_{\rm min}\sim0.01/2\pi$. 
The unfaithfulness computed from $Mf_{\rm min}\sim 0.0022/2\pi$
up to the merger is only weakly dependent on the
particular tidal approximant. 
However, tidal effects become significant at higher
frequencies, and if the $\bar{F}$ computations are restricted to higher
frequencies significant differences amongst the approximants emerge.
$\Psi_T^{\rm NRP}$ has the smallest unfaithfulness. 
For MS1b$_{1.35}^{0.00}$ (top panel), in particular, 
the proposed tidal approximant has an
unfaithfulness about one order of magnitude smaller than TaylorF2.  
For SLy$_{1.35}^{0.00}$ (bottom panel) the unfaithfulness is $\bar{F}<
0.03$ for all tidal approximants, indicating that the largest
contribution due to tidal effects comes from the strong-field--NR regime.
\\

\section{Conclusion}
The tidal approximants proposed here can be efficiently used for both
GW searches and parameter estimation of BNS events
and have shown their use for the interpretation of the recent BNS observation 
GW170817~\cite{TheLIGOScientific:2017qsa,Abbott:2018wiz,Abbott:2018exr,Dai:2018dca}~\footnote{
Details about the implementation in the LSC Algorithm Library Suite 
and extensive tests of the model can be found in~\cite{Dietrich:2018uni}.}.

For data-analysis
applications it is trivial to reparametrize the tidal coupling constant
$\kappa_{\rm eff}^T$ in terms of the mass ratio and (combinations of) the
dimensionless tidal parameters that are shown to be optimal for those
purposes \cite{Favata:2013rwa,Wade:2014vqa}.  
The approximant is valid up to the moment of merger frequency defined
by NR simulations in \cite{Bernuzzi:2014kca}. The latter references quantifies the
frequency corresponding to the peak of the waveform's amplitude in
term of the tidal polarizability coefficient $\kappa^T_2$; the
amplitude's peak formally marks the end of the chirp signal from BNS.
Although our work focused uniquely on the GW phase evolution, 
tidal corrections to the amplitude could also be added following
\cite{Damour:2012yf}. 
Future research will aim at improving the approximants using more NR
data and at including precession effects~\cite{Chatziioannou:2013dza,Hannam:2013oca}. 
Our work also proves that high-precision BNS simulations for GW
astronomy (similar to those used for the first BBH detections) are
now within reach of current technology.
%
%% Although we focused on BNSs, by setting $k_2^A=0$ and 
%% with the appropriate choice of the resulting
%% $\kappa_{\rm eff}^T$ one could construct waveforms for 
%% black-hole--neutron star binaries.
\\
%

%% ______________________________________________________________
\begin{acknowledgments}
\section{Acknowledgments} 
  It is a pleasure to thank S.~Babak, B.~Br\"ugmann, A.~Buonanno, R.~Cotesta, 
  W.~Del Pozzo, T.~Hinderer, B.~Lackey, K.~Kawaguchi, A.~Nagar, S.~Ossokine, and
  M.~P\"urrer for discussions and comments.
  We thank A.~Nagar and T.~Hinderer for providing the EOB waveforms. 
  We are grateful to E.~Poission for suggesting the definition of Eq.~\eqref{eq:kappa}.
  S.B. acknowledges support by the European Union's H2020 under ERC Starting Grant, grant
  agreement no. BinGraSp-714626. 
  W.T. was supported by the National Science Foundation under grant
  PHY-1305387.
  Computations were performed on SuperMUC at the LRZ (Munich) under 
  the project number pr48pu, Jureca (J\"ulich) 
  under the project number HPO21, Stampede 
  (Texas, XSEDE allocation - TG-PHY140019), 
  Marconi (ISCRA-B, under the project number HP10BMAB71), and 
  Marconi (PRACE, proposal number 2016153522).
\end{acknowledgments}

%%______________________________________________________________
%\bibliographystyle{revtex}     
\bibliography{paper20170609.bbl}     
%\bibliography{refs}     

\newpage

\onecolumngrid 

\section{SUPPLEMENTARY MATERIAL}

\subsection{Simulations overview}

The physical parameters of the BNS configurations and the grid
configurations used in the simulations are summarized in Tab.~\ref{tab:config}. 
The configurations consist of equal mass ($M_A=M_B$) 
binaries with aligned or anti-aligned spins ($\chi_A=\chi_B$). 
In total, nine (37) new configurations (simulations) 
have been performed for the scope of this paper. BNS configurations
are indicated by the EOS, the masses (subscript), and the spin
(superscript), i.e. ${\rm EOS}^{\chi_A}_{M_A}$.
We also use non-spinning data from the simulations described in 
\cite{Dietrich:2017feu} with the notation (${\rm EOS}_{M_A + M_B}$). 

Our initial configurations are constructed with the pseudospectral  
SGRID code~\cite{Tichy:2006qn,Tichy:2009yr,Tichy:2009zr,Dietrich:2015pxa}, 
which makes use of the constant rotational velocity 
approach~\cite{Tichy:2011gw,Tichy:2012rp} to construct constraint 
solved spinning BNS configurations in hydrodynamical equilibrium. 
Eccentricity reduced initial data are constructed 
following~\cite{Kyutoku:2014yba,Dietrich:2015pxa} by applying an 
iteration procedure varying the binary's initial 
radial velocity and the eccentricity parameter.

Evolutions are performed with the BAM
code~\cite{Brugmann:2008zz,Thierfelder:2011yi,Dietrich:2015iva,Bernuzzi:2016pie}.
We use 7 mesh refinement levels labeled with $l=0,...,6$
with grid spacing $h_l=h_0/2^{l}$ for $l>0$ and number of points per
direction $n_l$. 
Different grid resolutions named R1, R2, etc. have been employed; the NS diameter is typically covered with
$n_6=64,96,128,192,256$ points for respectively R1 to R5 (Table~\ref{tab:config}.)
The numerical fluxes for the general relativistic hydrodynamics 
are computed as in~\cite{Bernuzzi:2016pie} based on a 
flux-splitting approach with the local Lax-Friedrich flux
and after reconstruction of the characteristic
fields~\cite{Jiang:1996,Suresh:1997}. 
This method guarantees clear convergence and a 
robust error assessment of the numerical 
simulations, see below. 

\begin{table*}[h]
  \centering    
  \caption{BNS configurations. 
    The first column defines the name of the configuration
    with the notation: EOS$_{M^A}^{\chi^A}$.
    The subsequent columns describe the 
    corresponding physical properties of the individual stars: 
    the EOS~\cite{Read:2008iy},
    the gravitational mass of the individual stars $M_{A,B}$, 
    the binary mass $M=M_A+M_B$, 
    the binary's baryonic mass $M_{b}$, 
    the stars' dimensionless spin $\chi_{A,B}$, 
    the effective spin $\chi_{\rm eff}$,
    the dimensionless quadrupolar tidal coupling constant $\kappa^T_2$,
    the initial dimensionless GW frequency $M \omega_{22}(0)$, 
    the ADM mass $M_{\rm ADM}(0)$ and 
    angular momentum $J_{\rm ADM}(0)$ and 
    residual eccentricity estimated from the 
    proper distance, see~\cite{Dietrich:2015pxa}. 
    The last columns indicate the resolutions employed for BAM's
    evolution grid. We use 7 mesh refinement levels $l=0,...,6$
    with refinement ratio 2:1 and grid spacing $h_l=h_0/2^{l}$ for $l>0$.
    We report the number of points per
    direction in the finest level, $n_6$, for all the runs and the
    grid resolution $h_6$ in the finest level of the most resolved
    run. Grid resolutions of other runs can be obtained using $h'=h\cdot
    n/n'$. }
%  \begin{tabular}{l||ccccc|ccccc|c}       
\begin{small}
\begin{tabular}{l||ccccccccccc|cc}        
\hline
\hline
  Name & EOS & $M_{A,B}$ & $M$ & $M_b$ & $\chi_{A,B}$ & $\chi_{\rm
     eff}$ & $\kappa^T_2$ & $M \omega_{22}(0)$ &$ M_{\rm ADM}(0)$ &
     $J_{\rm ADM}(0)$ & $e [10^{-3}]$ & $n_6$ & $h_6$  \\
     \hline
MS1b$_{1.35}^{-0.10}$ & MS1b & 1.3504 & 2.7008 & 2.9351 & -0.099 & -0.082 & 288.0 & 0.0357 & 2.6795 & 7.4858 & 1.8 & (64,96,128,192) & 0.097 \\
MS1b$_{1.35}^{0.00}$  & MS1b & 1.3500 & 2.7000 & 2.9351 & +0.000 & +0.000 & 288.0 & 0.0357 & 2.6786 & 7.8021 & 1.7 & (64,96,128,192) & 0.097 \\
MS1b$_{1.35}^{0.10}$  & MS1b & 1.3504 & 2.7008 & 2.9351 & +0.099 & +0.082 & 288.0 & 0.0357 & 2.6793 & 8.1292 & 1.9 & (64,96,128,192) & 0.097\\
MS1b$_{1.35}^{0.15}$  & MS1b & 1.3509 & 2.7018 & 2.9351 & +0.149 & +0.123 & 288.0 & 0.0357 & 2.6802 & 8.3054 & 1.8 & (64,96,128,192)  & 0.097\\
\hline
 H4$_{1.37}^{0.00}$   & H4   & 1.3717 & 2.7435 & 2.9892 & +0.000 & +0.000 & 190.0 & 0.0367 & 2.7213 & 8.0052 & 0.9 & (64,96,128,192) & 0.083 \\
 H4$_{1.37}^{0.14}$   & H4   & 1.3726 & 2.7452 & 2.9892 & +0.141 & +0.117 & 190.0 & 0.0368 & 2.7229 & 8.4897 & 0.4 & (64,96,128,192) & 0.083\\
\hline
SLy$_{1.35}^{0.00}$   & SLy  & 1.3500 & 2.7000 & 2.9892 & +0.000 & +0.000 & 73.5 & 0.0379 & 2.6778 & 7.6860 & 0.4 & (64,96,128,192,256) & 0.059\\     
SLy$_{1.35}^{0.05}$   & SLy  & 1.3502 & 2.7003 & 2.9892 & +0.052 & +0.043 & 73.5 & 0.0379 & 2.6780 & 7.8588 & 0.4 & (64,96,128,192) & 0.078\\     
SLy$_{1.35}^{0.11}$   & SLy  & 1.3506 & 2.7012 & 2.9892 & +0.106 & +0.088 & 73.5 & 0.0379 & 2.6789 & 8.0391 & 0.7 & 64,96,128,192 & 0.078\\ 
\hline      
\hline
     \end{tabular}
     \end{small}
 \label{tab:config}
\end{table*}

\subsection{Simulations accuracy}

\subsubsection{Eccentricity}

\begin{figure*}
\centering
\begin{minipage}[t]{.48\textwidth}
   \includegraphics[width=\textwidth]{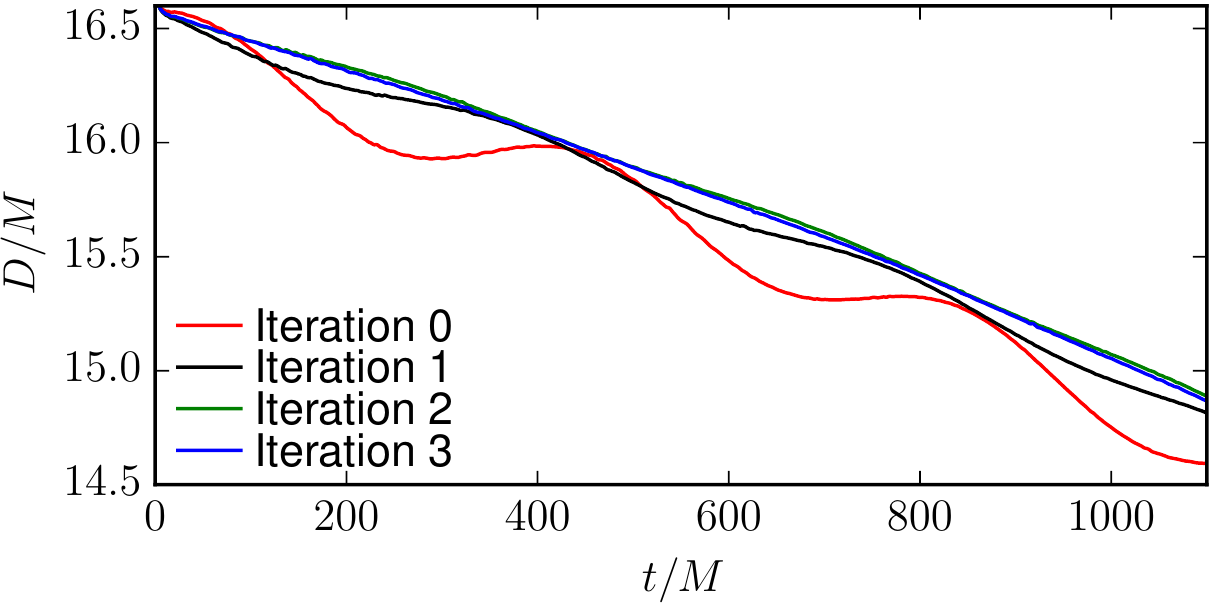}
   \caption{ \label{fig:ecc-red}
   Proper distance along the connection line between the two NSs centers for SLy$_{1.35}^{0.00}$.  
   }
\end{minipage}\qquad
\begin{minipage}[t]{.48\textwidth}
   \includegraphics[width=\textwidth]{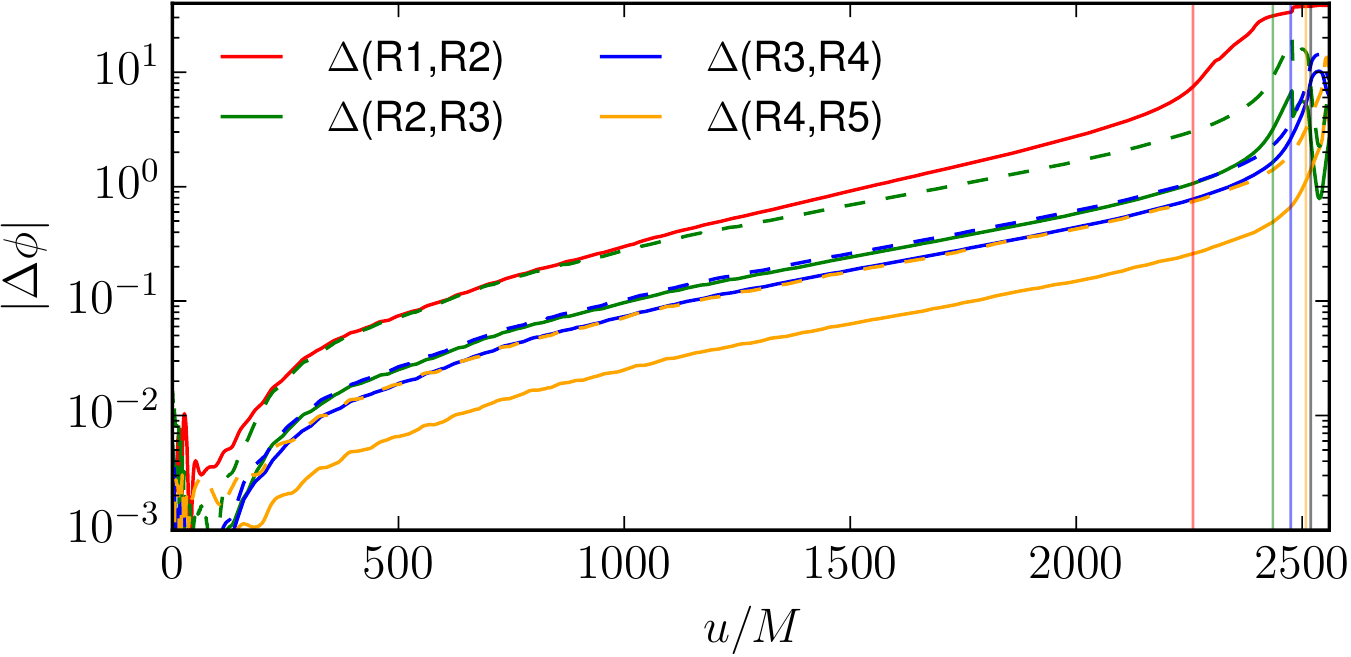}
   \caption{ \label{fig:convergence_phase}
   Phase difference for setup SLy$_{1.35}^{0.00}$ (solid lines).
   Dashed lines are rescaled phase differences assuming 
   second order convergence, which is achieved 
   for resolution R2 and above. Straight vertical 
   lines mark the moment of merger, i.e.~the peak 
   in the GW amplitude. 
   }
\end{minipage}
\end{figure*}

We construct eccentricity reduced initial data by means of 
an iterative procedure that monitors and varies the binary's initial 
radial velocity and the eccentricity parameter, see Eq.~(2.37)
and (2.39) of~\cite{Dietrich:2015pxa}.
As an initial guess, quasi-equilibrium configurations 
in the usual quasi-circular orbit are employed for which 
residual eccentricities are in the range of $e\sim10^{-2}$. 
The steps of the iterative procedure are then, 
(i) evolve the data for $\sim3$ orbits, 
(ii) measure the eccentricity $e$, 
   for which we use the proper distance as described in~\cite{Dietrich:2015pxa}, and 
(iii) re-compute the initial data with adjusted parameters. 
As an exemplary case, the iteration procedure for the SLy$_{1.35}^{0.00}$ case 
is presented in Fig.~\ref{fig:ecc-red}. 
Target residual eccentricities $e \sim10^{-3}$ 
are usually achieved within three iterations~\cite{Dietrich:2015pxa}.

\subsubsection{Waveform's error-budget}

The GW metric multipoles 
\begin{equation}
r\, h_{\ell m} (t) = A_{\ell m}(t) e^{- i \phi_{\ell m}(t)}  
\end{equation}
are constructed from the curvature multipoles using 
frequency domain integration of Ref.~\cite{Reisswig:2010di}. 
In this work only the dominant (2,2)-mode is considered and 
indices are dropped in the following. The retarded time is defined as 
$u=t-r_*$, where $r_*(M)$ is the
tortoise coordinate of the Schwarzschild spacetime of mass $M$ computed from the
coordinate (isotropic) radius at which GWs are extracted (see below).

Uncertainties due to truncation errors are estimated
following \cite{Bernuzzi:2016pie}. 
As an exemplary case we present the phase difference between 
different resolutions for SLy$^{0.00}_{1.35}$ in
Fig.~\ref{fig:convergence_phase}.  
Second order convergence (dashed lines) is achieved for resolutions R2
and higher.  
For a better approximation of the waveform 
we follow the description of~\cite{Hannam:2007ik} and
apply a Richardson extrapolation for the phase using the highest three 
available resolutions for each dataset. 

GWs are extracted at finite radii, where we pick $r=1000$ for our analysis. 
The numerical error introduced by finite radii extraction of the GW is 
obtained by comparing finite radii waveforms with 
second oder polynomial extrapolated waveform 
(similar results are obtained by including next-to-leading order terms, 
see~\cite{Lousto:2010qx,Bernuzzi:2016pie}).

\subsection{Time-domain approximants}

\subsubsection{Time-domain fit}

To obtain the time-domain fit, we split the interval $\mo \in I=[0,0.17]$ into three 
different intervals: $I_{\rm T2} = [0,0.0074]$, $I_{\rm EOB} = [0.0074,0.04]$, 
$I_{\rm NR} = [0.04,0.17]$. 
In $I_{\rm T2}$ we evaluate Eq.~\eqref{eq:T2} at $10000$ equally spaced points. 
For interval $I_{\rm EOB}$ we compute three tidal EOB waveforms~\cite{Bernuzzi:2014owa}
corresponding to the three irrotational runs of Tab.~\ref{tab:config}
and with starting frequency $\mo(0) = 0.0065$. We compute $\Delta \phi_T/\kappa_2^T$
by taking the difference of $\phi(\mo)$ for different BNS configurations. 
From the obtained curves we compute the average 
and interpolate on an equally spaced grid with a total of $5000$ grid points in $I_{\rm EOB}$. 
A phase shift is applied to the EOB data by minimizing the phase
difference between the T2 and EOB data in the interval
$\hat{\omega}\in[0.00715,00765]$. 
For the interval $I_{\rm NR}$ we compute $\phi_T/\kappa_2^T$ by taking the difference
between the irrotational NR data [as for the EOB waveforms]; we then
take the average of all obtained results, interpolate on an equally
spaced  grid with $500$ grid points, and fix the initial phase by minimizing the phase difference in 
$[0.04,0.044]$.

The obtained data on $I=I_{\rm T2}\cup I_{\rm EOB}\cup I_{\rm NR}$ are
fittd as a function of $x$, after factoring out the leading order
Newtonian term. The result is given in Eq.~\eqref{eq:fitT}; the high
frequency part of the fit is shown in the bottom panel of
Fig.~\ref{fig:Phiomega}.

\subsubsection{Waveform comparison}

\begin{figure*}[t]
   \includegraphics[width=1\textwidth]{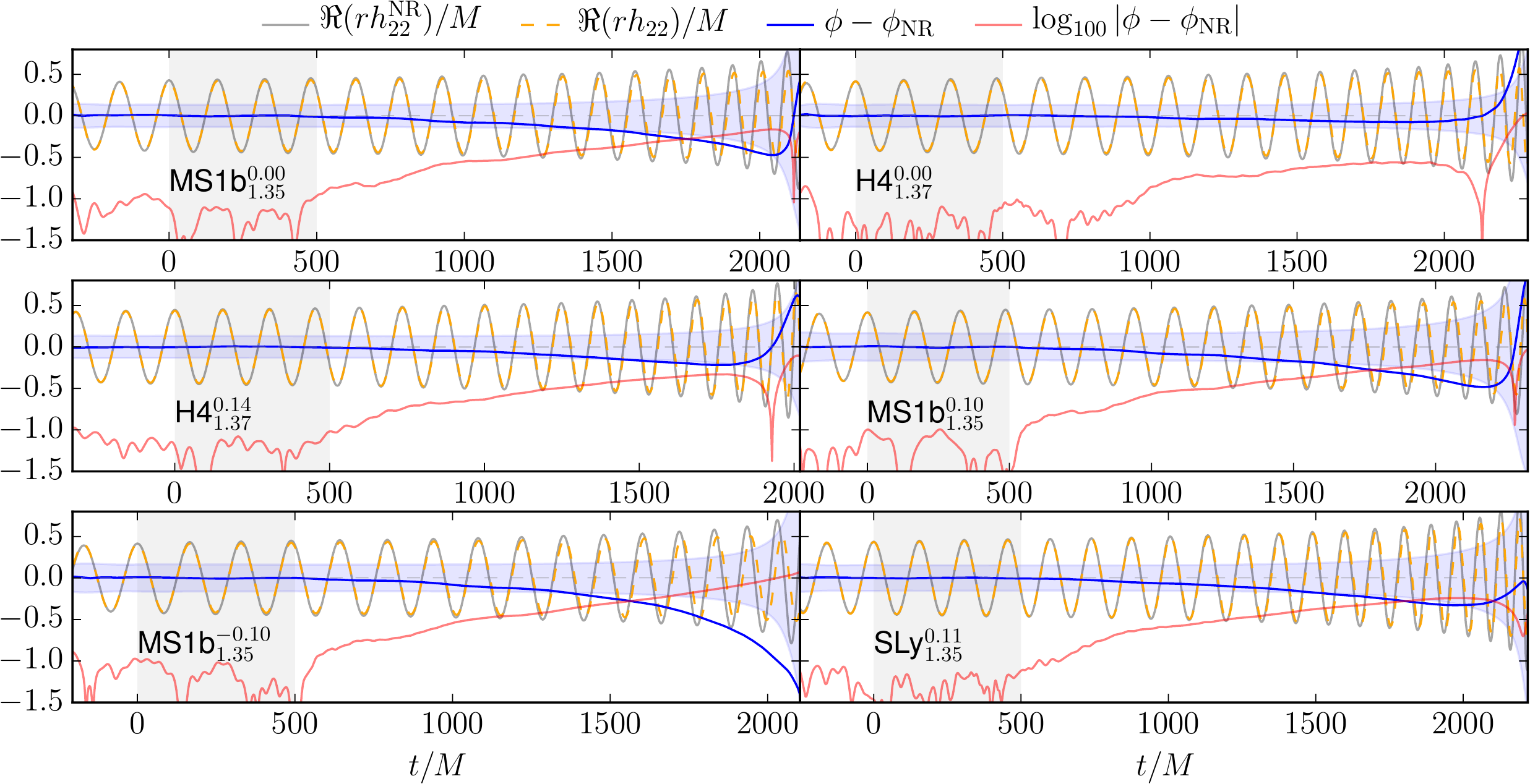}
   \caption{ \label{fig:performance_newspin} 
   GWs for different setups, see Tab.~\ref{tab:config}. 
   We compare our waveform model with full NR simulations. 
   For each configuration we show the real part obtained with Eq.~\eqref{eq:fitT} (orange) 
   and the real part of the comparison waveform (gray). 
   The dephasing between the model and the comparison waveform $\phi-\phi_{\rm NR}$ 
   is shown blue and $\log_{100}|\phi-\phi_{\rm NR}|$ is shown red.  
   The numerical uncertainty is shown as a blue shaded region and 
   the alignment region as a gray shaded interval. \
   }
 \end{figure*}

\begin{figure*}[t]
   \includegraphics[width=1\textwidth]{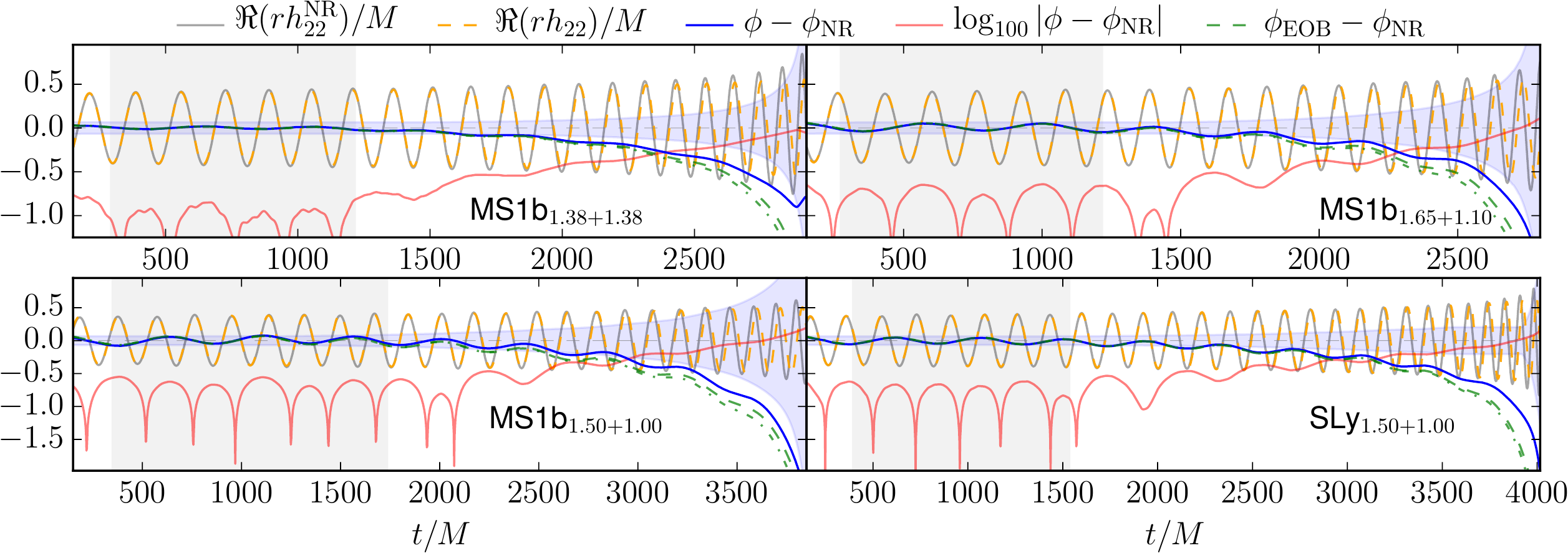}
   \caption{ \label{fig:performance_q} 
   Comparison of the proposed waveform model with NR simulations of~\cite{Dietrich:2017feu}. 
   For each configuration we show the real part obtained with~\eqref{eq:fitT} (orange) 
   and the real part of the NR waveform (gray). 
   The dephasing between the model and the NR data $\phi-\phi_{\rm NR}$ 
   is shown blue and $\log_{100}|\phi-\phi_{\rm NR}|$ is shown red. 
   We also include the phase between the NR data with respect to 
   the EOB models of~\cite{Hinderer:2016eia} (green dashed) 
   and~\cite{Bernuzzi:2014owa} (green dot-dashed).
   The numerical uncertainty is shown as a blue shaded region and 
   the alignment region as a gray shaded interval.
   Considered configurations are: 
   an equal mass setups for the MS1b with masses $M_A=M_B=1.375$ (upper left); 
   an unequal mass setups setup with $M_A=1.65,M_B=1.10$ for MS1b (upper right), and 
   unequal mass setups with $M_A=1.50,M_B=1.00$ for MS1b (lower left) and 
   SLy (lower right) (bottom row).
   }
 \end{figure*}

\begin{figure*}[t]
   \includegraphics[width=1\textwidth]{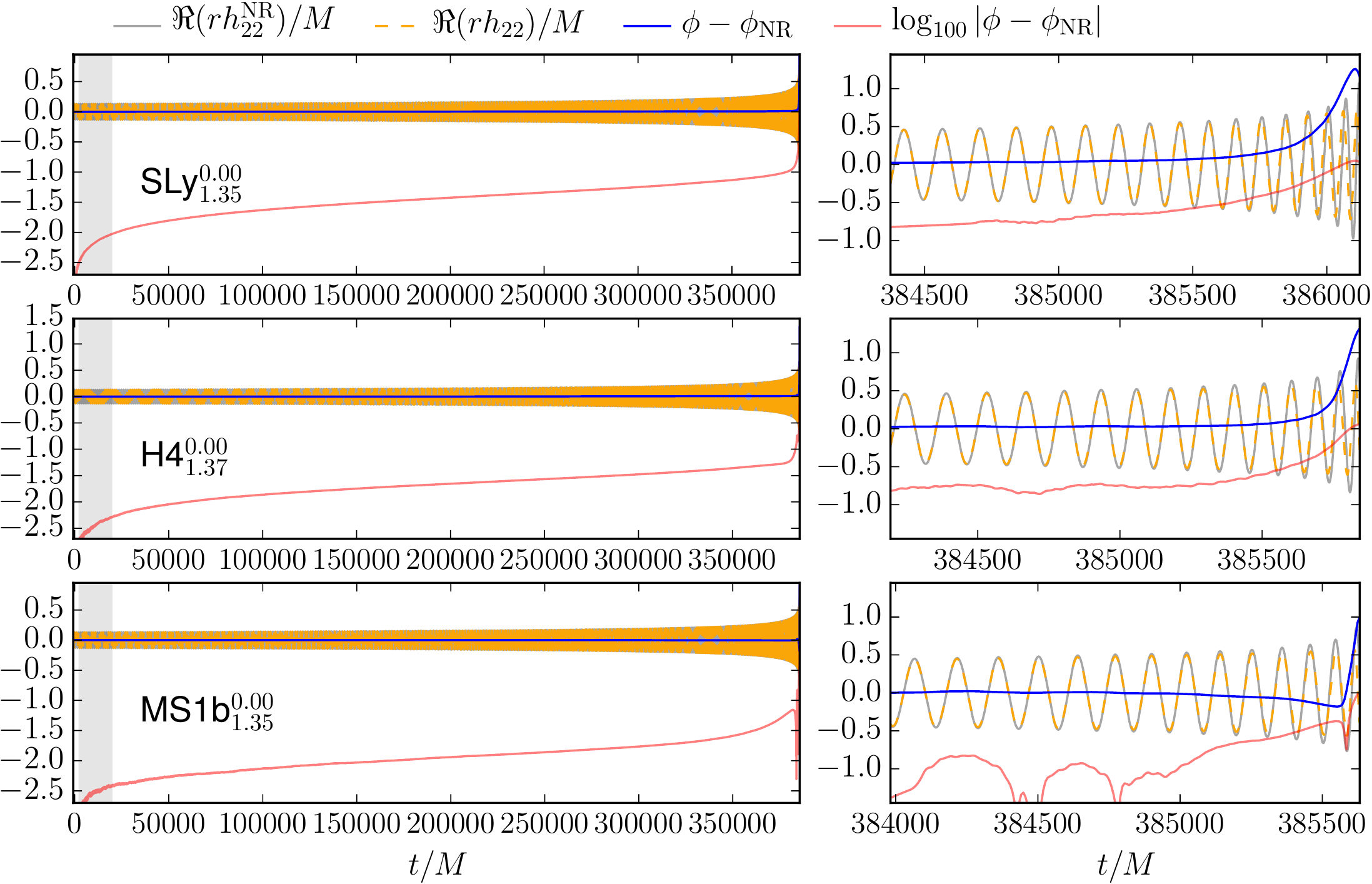}
   \caption{ \label{fig:performance_EOB} 
   We compare our waveform model with hybrid NR-tidal EOB waveforms.
   For each configuration we show the real part obtained with~\eqref{eq:fitT} (orange) 
   and the real part of the hybrid waveform (gray). 
   The dephasing between the model and the hybrid $\phi-\phi_{\rm NREOB}$ 
   is shown blue and $\log_{100}|\phi-\phi_{\rm NREOB}|$ is shown red.  
   The alignment region is marked as a gray shaded interval. 
   The simulations cover about 300 orbits before the merger.
   }
 \end{figure*}

In addition to the exemplary cases in the main text, we further test the
performances of the prosed model comparing to 
\begin{itemize}
\item Waveforms from all the simulations listed in Tab.~\ref{tab:config}
[See Fig.~\ref{fig:performance_newspin} for examples]. 
\item Equal and unequal mass NR waveforms of
Ref.~\cite{Dietrich:2017feu} [see Fig.~\ref{fig:performance_q} in which we also include tidal EOB waveforms].
\item Hybrid EOB-NR waveforms with a starting frequency of 75 Hz.
The hybrid waveforms are constructed by combining tidal EOB 
waveforms of~\cite{Bernuzzi:2014owa} with the highest resolution 
irrotational data presented in Tab.~\ref{tab:config} [See Fig.~\ref{fig:performance_EOB}
for the results]. 
\end{itemize}
We find that our time-domain approximant is robust for a variation of
the binary parameters and the waveform length. In particular, spurious
effects due to time-domain waveform alignment do not influence 
significantly our results.

\subsection{Frequency-domain approximants}

The 2.5PN TaylorF2 expression of Damour et al.~\cite{Damour:2012yf}
(DNV) with which we compare reads for equal mass BNSs:
      \begin{eqnarray}
\label  {eq:Psi2.5PN}
  \Psi_T^{\rm 2.5PN}& =  
  \kappa^T_2 \ct_{\rm Newt} x^{5/2} ( 1 + \ct_1 x + \ct_{3/2}x^{3/2} +
  \ct_{2}x^{2}  + \ct_{5/2}x^{5/2} ) \ ,
\end{eqnarray}
where 
\begin{eqnarray}
\ct_{\rm Newt} = -\frac{39}{4}, \quad 
\ct_{1} = \frac{3115}{1248}, \quad 
\ct_{3/2} = -\pi, \quad  
\ct_{2} = \frac{23073805}{3302208} + \frac{20}{81} \bar{\alpha}^2_2+
\frac{20}{351}\beta^{22}_2, \quad 
\ct_{5/2} = -\pi \frac{4283}{1092}\ .
\end{eqnarray}
For equal masses $\bar{\alpha}^2_2=85/14$.
The expression above includes tail terms up to 2.5PN order and the 2PN
is computed up to an unknown (not yet
calculated) coefficient which we set $\beta^{22}_2=0$.

To validate the frequency approximant we compute the {\it mismatch}
(or {\it unfaithfulness})
\begin{equation}
\bar{F} = 1 - \max_{\phi_c,t_c} \frac{(h_1(\phi_c,t_c)|h_2)}{\sqrt{(h_1|h_1),(h_2,h_2)}}
\end{equation}
with $\phi_c,t_c$ an arbitrary phase and time shift, and the
noise-weighted overlap defined as  
\begin{equation}
 (h_1,h_2) = 4 \Re \int_{f_{\rm min}}^{f_{\rm
 max}} \frac{\tilde{h}_1(f) \tilde{h}_2(f)}{S_n(f)} \text{d} f \ . 
\end{equation}
Above, $S_n(f)$ is the one-sided power spectral density of the detector
noise, where we use the \verb#ZERO_DET_high_P# noise curve
of~\cite{Sn:advLIGO}. 
The value of $\bar{F}$ indicates the loss in signal-to-noise ratio
(squared) when the waveforms are aligned in time and phase. 
Template banks are constructed so that the maximum value is
$\max(\bar{F})=0.03$. Such mismatch corresponds to a maximum loss in 
event-rate of $\sim0.09$.

%%______________________________________________________________

%\bibliographystyle{revtex}     
%%\bibliography{refs} 

\end{document}